\documentclass[aps,pre,twocolumn,superscriptaddress,amsmath,showpacs]{revtex4}

\usepackage{graphicx, enumitem, multirow}
\usepackage[outercaption]{sidecap}
\usepackage{hyperref}
\newcommand{\footstar}[1]{$^*$ \footnotetext{$^*$#1}}
\begin{document}

\title{Extended finite-size scaling of synchronized coupled oscillators}

\author{Chulho Choi}
\affiliation{Department of Physics and Astronomy, Seoul National University, Seoul 151-747, Korea}

\author{Meesoon Ha}
\email[Corresponding author: ]{msha@chosun.ac.kr}
\affiliation{Department of Physics Education, Chosun University,
Gwangju 501-759, Korea}

\author{Byungnam Kahng}
\affiliation{Department of Physics and Astronomy, Seoul National University, Seoul 151-747, Korea}

\date{\today}

\begin{abstract}
We present a systematic analysis of dynamic scaling in the time
evolution of the phase order parameter for coupled oscillators
with non-identical natural frequencies in terms of the Kuramoto
model. This provides a comprehensive view of phase
synchronization. In particular, we extend finite-size scaling
(FSS) in the steady state to dynamics, determine critical
exponents, and find the critical coupling strength. The dynamic
scaling approach enables us to measure not only the FSS exponent
associated with the correlation volume in finite systems but also
thermodynamic critical exponents. Based on the extended FSS
theory, we also discuss how the sampling of natural frequencies
and thermal noise affect dynamic scaling, which is numerically
confirmed.
\end{abstract}



\maketitle

\begin{SCfigure*}[]
\includegraphics[width=0.29\textwidth]{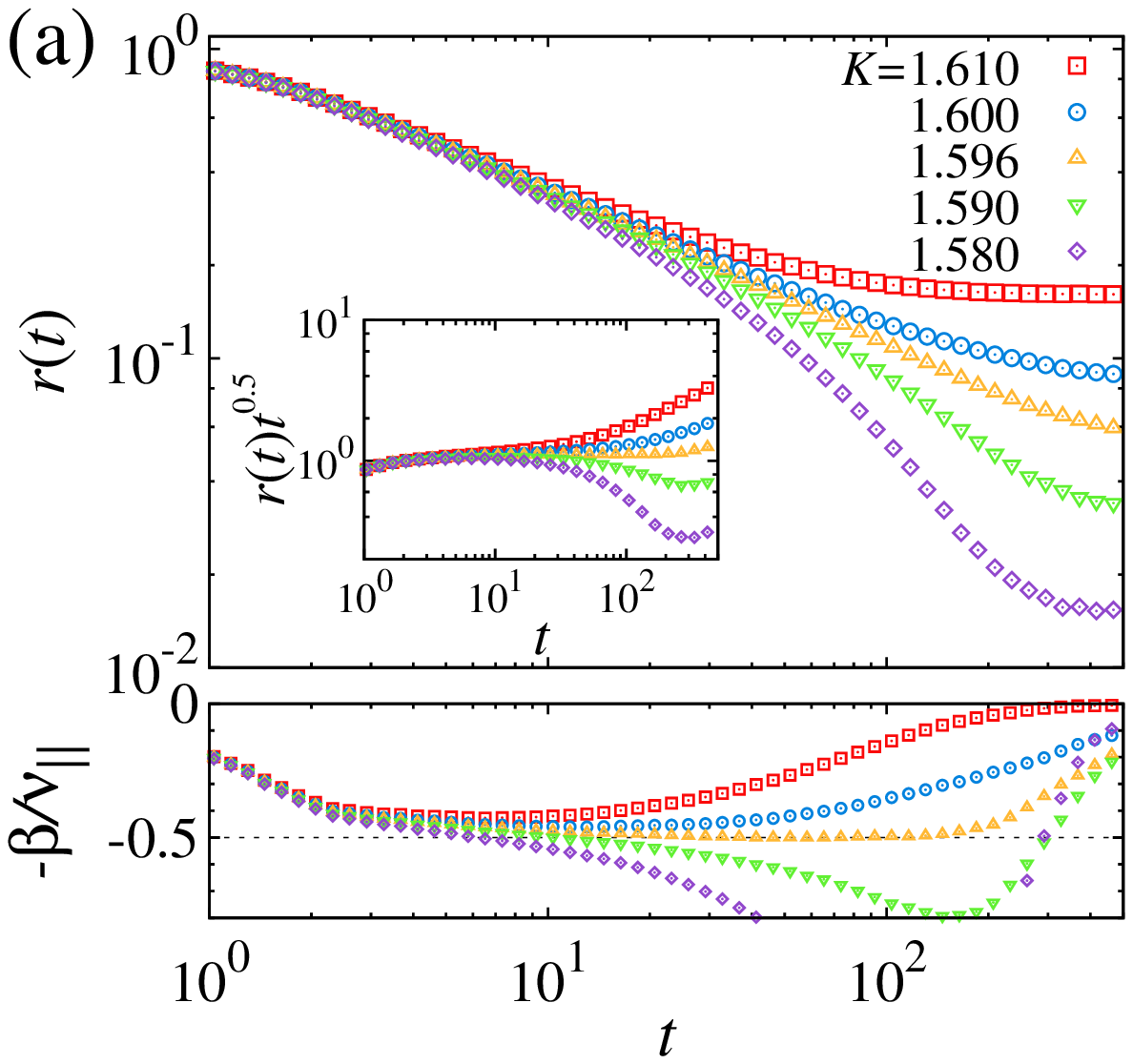}
\includegraphics[width=0.29\textwidth]{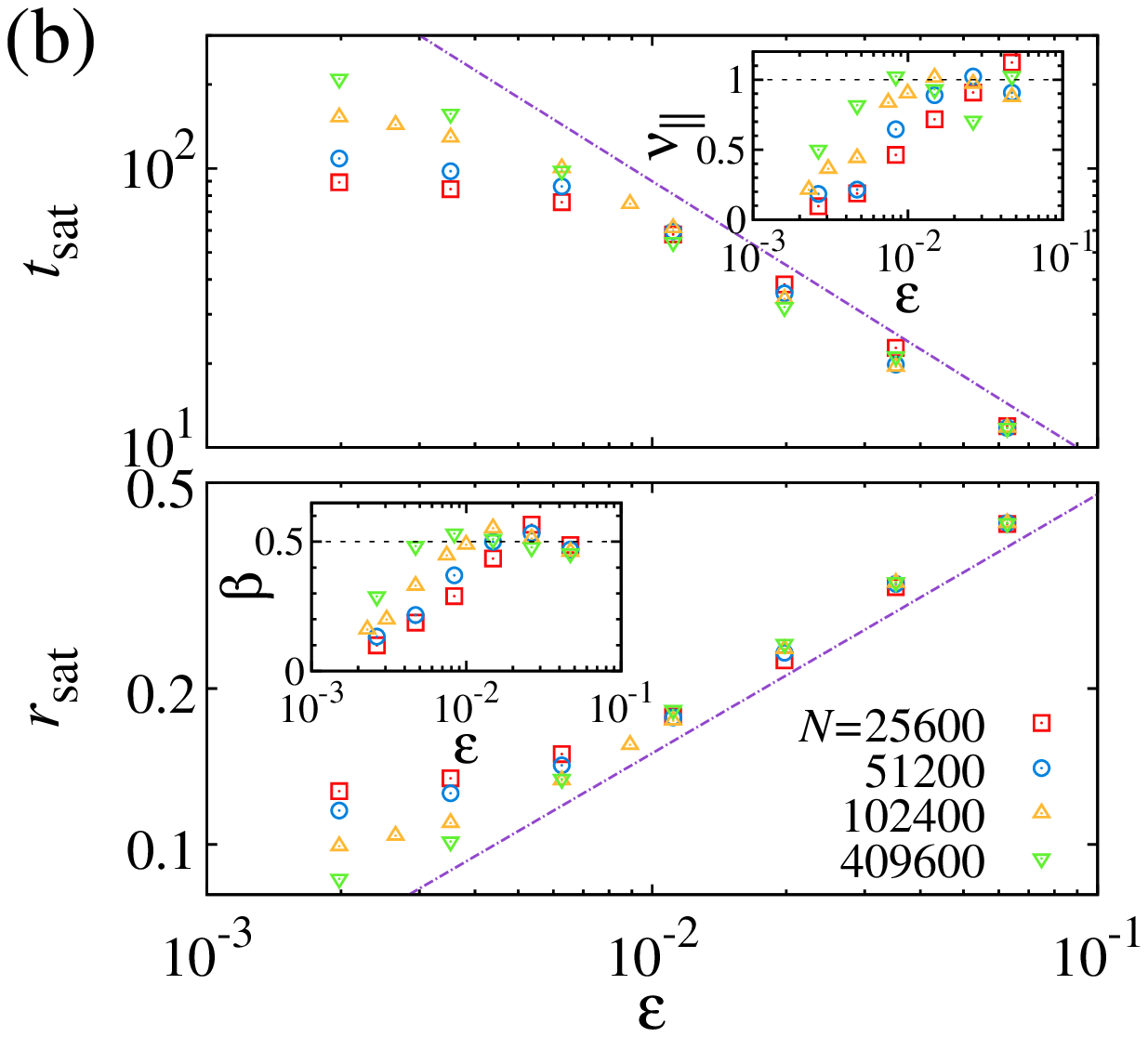}
\caption{(Color online) (a) Temporal behaviors of
$r_{\downarrow}(t)$ near the criticality [$K_{\rm
c}=\sqrt{8/\pi}\simeq 1.596$ for Gaussian $g(\omega)$] at
$N=819200$. Effective exponent plots in the lower panel indicate
the value of $K_{\rm c}$ where $r(t)\sim
t^{-\beta/\nu_{\parallel}}$ with $\beta/\nu_{\parallel}=1/2$. (b)
Based on critical behaviors $t_{\rm sat}$ and $r_{\rm sat}$ near
$\epsilon=0$, two thermodynamic exponents
($\nu_{\parallel}=1,~\beta=1/2$) are measured (insets) as $N$
increases. Here data are obtained from the random sampling of
$\{\omega_j\}$ and $r(0)=1$ (at least 200 ensembles).}
\label{fig:off-scaling}
\end{SCfigure*}
\begin{figure*}[]
\centering
\includegraphics[width=0.3\textwidth]{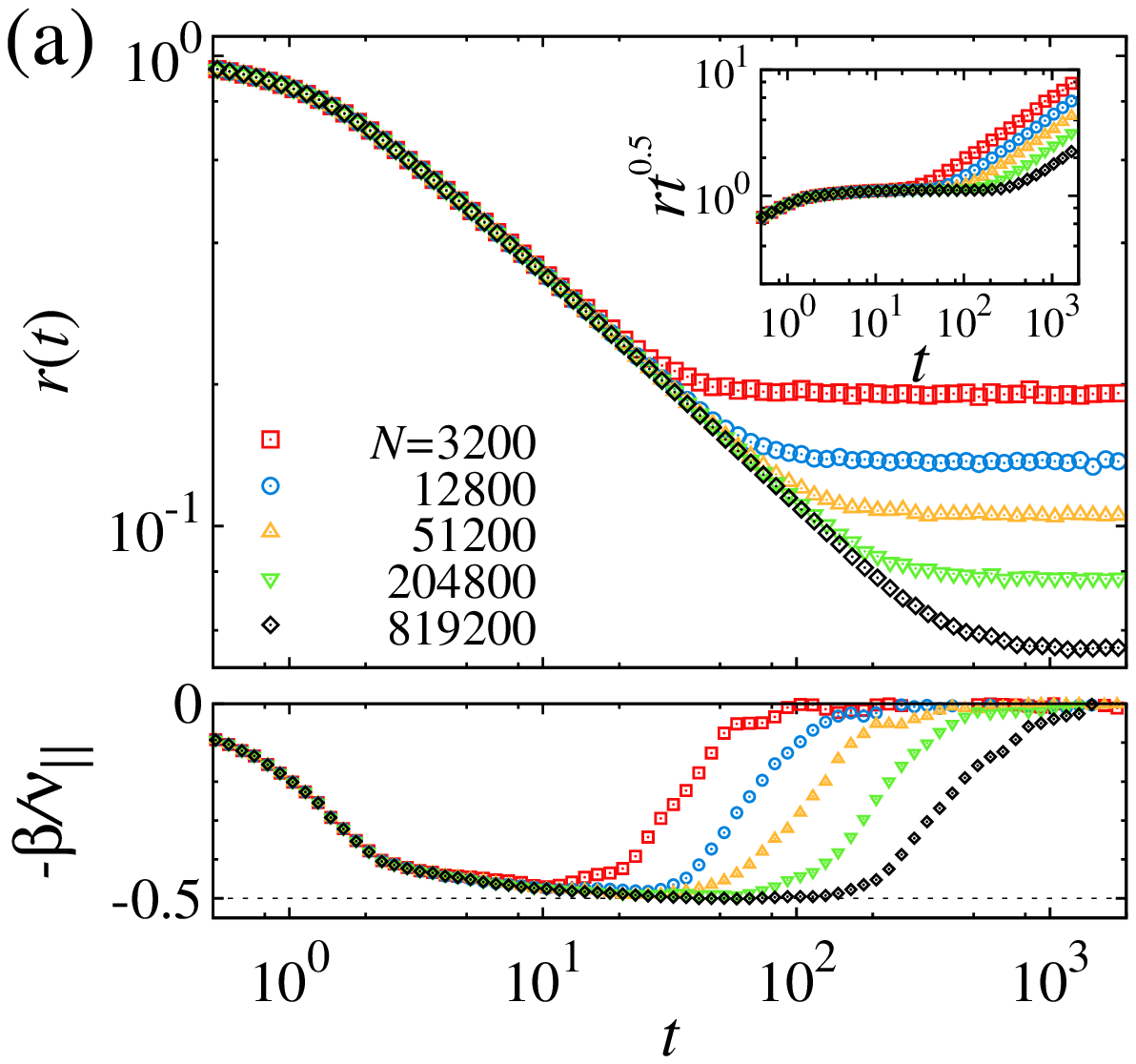}
\includegraphics[width=0.3\textwidth]{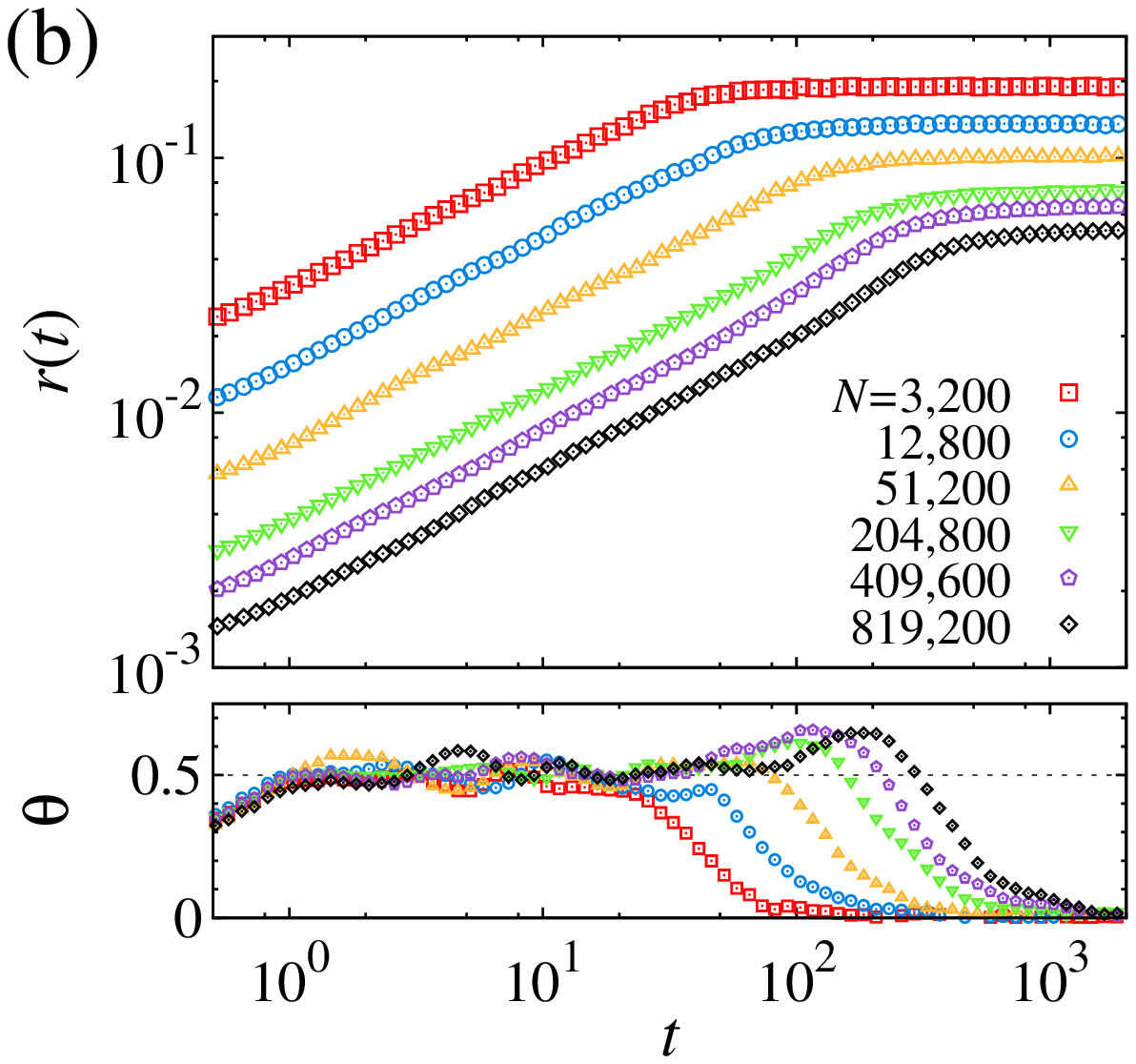}
\includegraphics[width=0.3\textwidth]{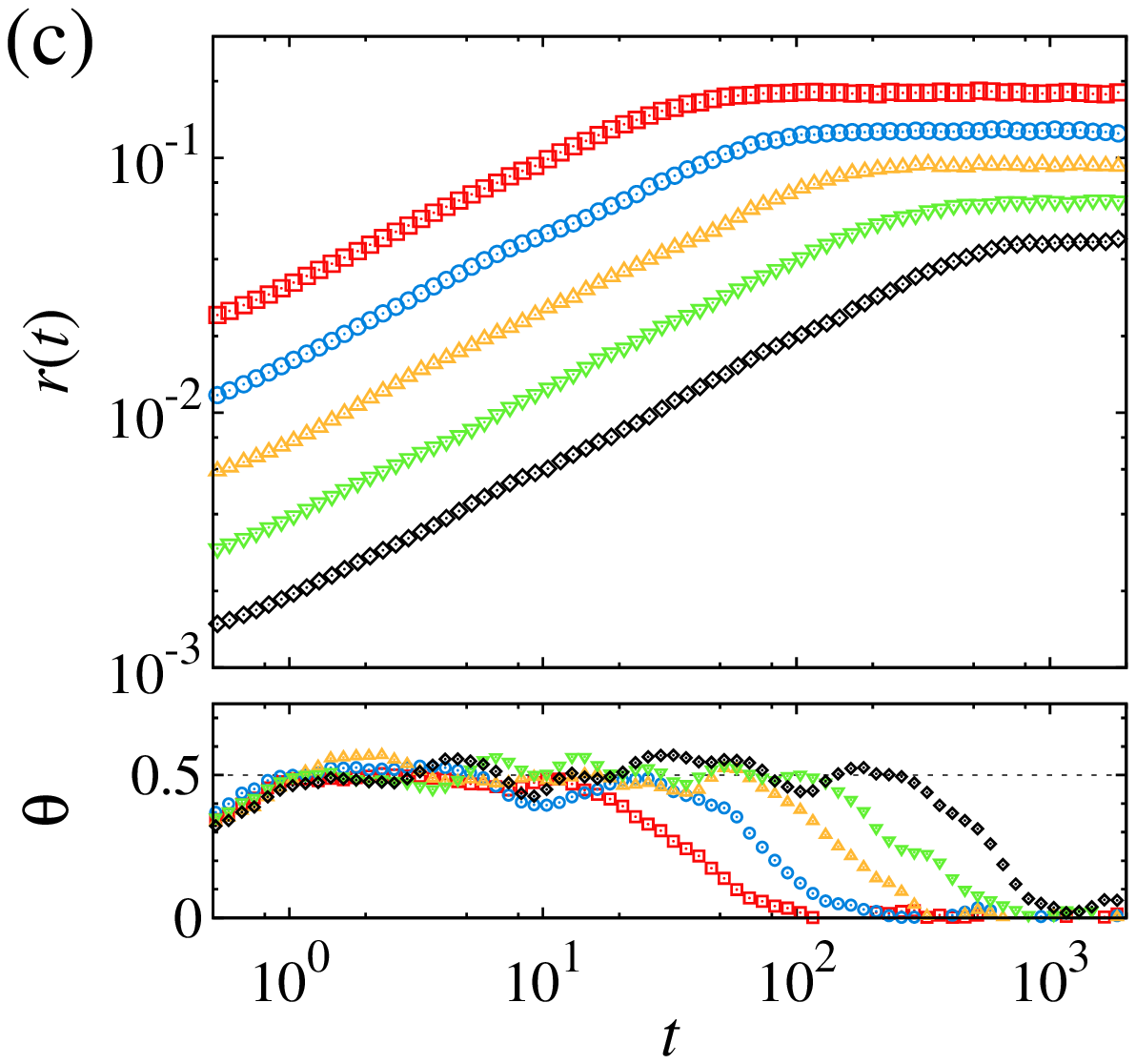}
\caption{(Color online) For the random sampling of $\{\omega_j\}$,
$r(t)$ is plotted at $K=K_{\rm c}(T)$ at various $N$ with the
corresponding effective exponent
($\beta/\nu_{\parallel}\mbox{~or~}\theta$) plots: When the KM
starts (a) at a coherent [$r(0)= 1$] with
$\beta/\nu_{\parallel}=1/2$; (b) at an incoherent state [$r(0)\sim
N^{-1/2}$] with $\theta=1/2~\to~3/4$; (c) at the same state as (b)
but containing thermal noise ($T=0.1$) with $\theta=1/2$. Note
that the same symbol (color) corresponds to the same size as
described in (b) unless any other explanations are provided.}
\label{fig:DS4Gaussian-}
\end{figure*}
\begin{figure}[]
\centering
\includegraphics[width=0.625\columnwidth]{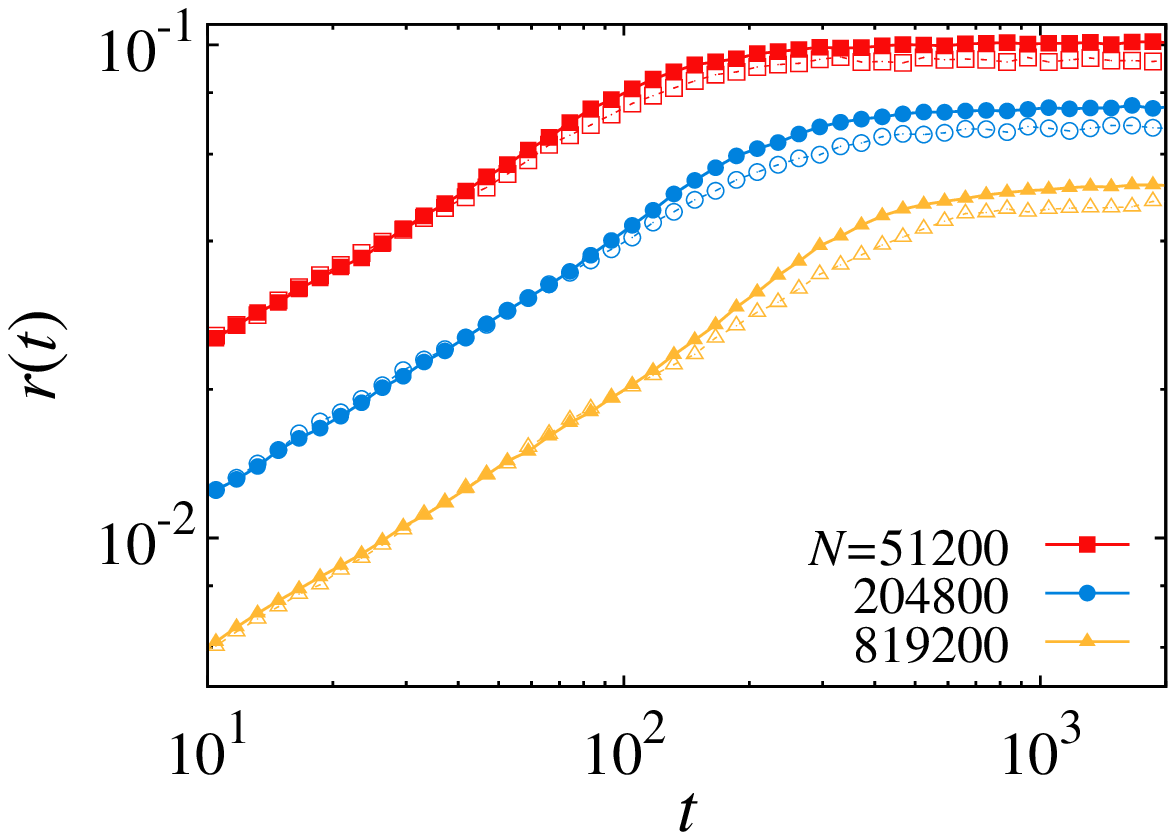}\\
\includegraphics[width=0.625\columnwidth]{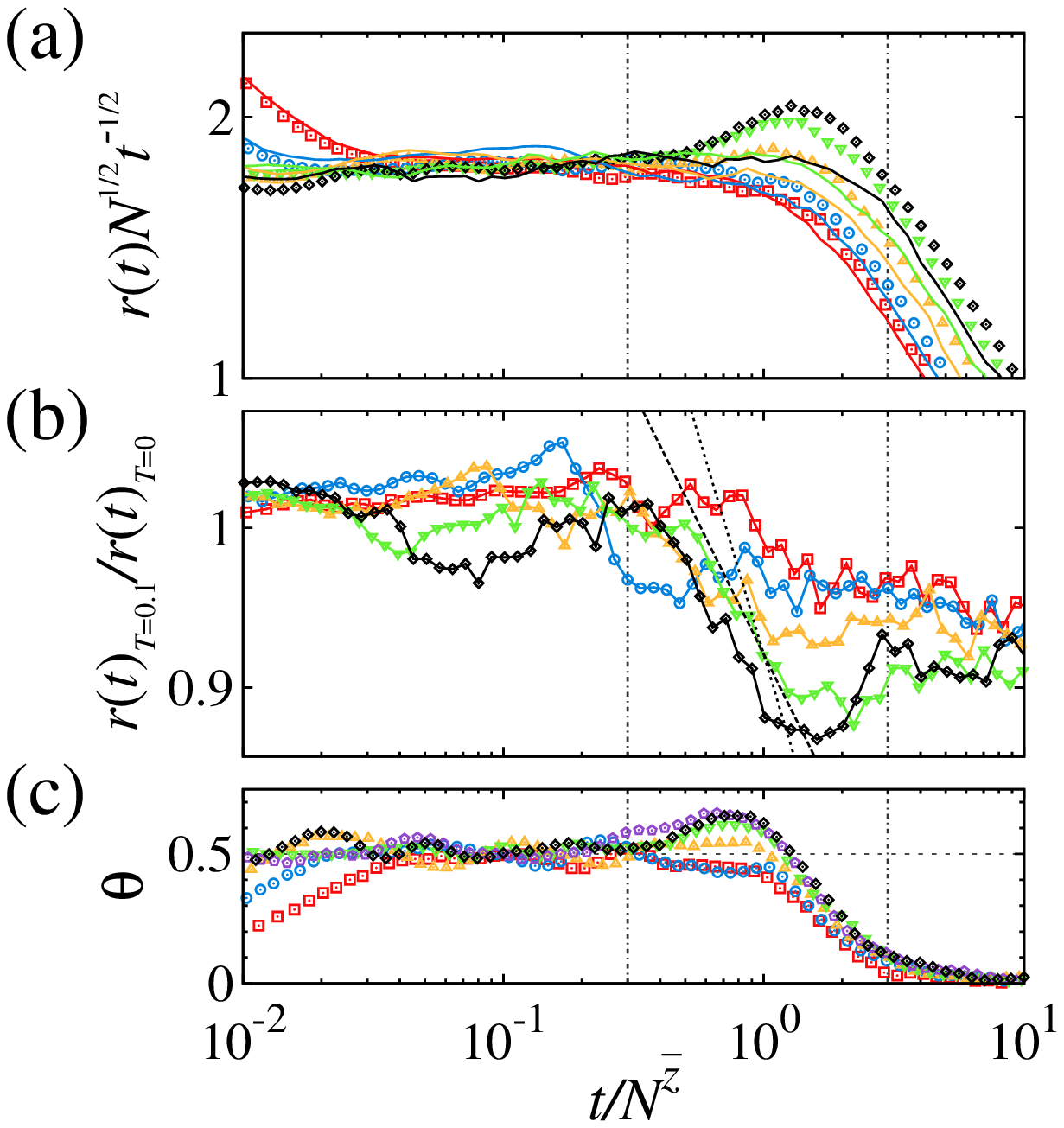}
\caption{(Color online) The effect of thermal noise on dynamic
scaling of $r(t)$ at $K=K_{\rm c}(T)$ is tested for two cases,
$T=0.1$ (noisy, open symbols) and $T=0$ (noiseless, filled
symbols), at three different $N$ values. (a) The noisy random
sampling case (lines for $T=0.1$) is compared to the noiseless
case (symbols for $T=0$). (b) The ratio of two cases are plotted.
Two straight lines are guides for eyes, of which slopes are -0.15
and -0.25, respectively. Based on our conjecture, it should be the
same as $\theta_{T=0.1}-\theta_{T=0}=-1/4$. (c) Scaling collapse
of effective exponents in the lower panel of
Fig.~\ref{fig:DS4Gaussian-}~(b) implies $t_{\rm cross}\sim t_{\rm
sat}\sim N^{\bar{z}}$ with $\bar{z}=2/5$. Here we use the same
data as those in the upper panel of
Fig.~\ref{fig:DS4Gaussian-}~(b) and (c).} \label{fig:comparison}
\end{figure}
\section{Introduction}
\label{intro}

Collective synchronization of coupled oscillators is a fascinating
phenomenon, where non-identical oscillators are spontaneously
coherent at the same frequency with identical phase angles with
each cycle (or a repeating sequence of phase angles over
consecutive cycles) with diverging scales. This cooperative
behavior is ubiquitous in real systems from well-known examples,
Josephson junction arrays, chemical oscillators and flashing of
fireflies~\cite{old-intro+review}, to most recent examples, such
as power
grids~\cite{Rohden2012PRL+Doefler2012PNAS+Motter2013NPHYS},
chimera states in oscillator networks~\cite{Martens2013PNAS}, and
neural networks~\cite{Takahashi2010PNAS+Kveraga2011PNAS}.

From theoretical point of view, such a remarkable phenomenon has
also become a central issue as an universal concept in nonlinear
science~\cite{books}. Kuramoto introduced a mathematically
tractable model of coupled nonlinear oscillators~\cite{Kuramoto}
as refining the earlier model by Winfree~\cite{Winfree}. Since
then, the Kuramoto model (KM) has played a role as the
paradigmatic model of synchronization. The KM is simple but
exhibits rich behaviors; among them, the synchronization
transition is one of fundamental problems. At the transition,
oscillators' phases are tuned by the critical coupling strength
against non-identical natural frequencies, and eventually reach a
phase-locked state (frequency entrainment) including in-phase
synchronization with exactly the same value.

A continuous synchronization transition in the KM was firstly
characterized in the mean-field (MF) picture, and accomplished by
solving a self-consistent equation of the order parameter. The MF
solution of critical exponents associated with the order parameter
($r\sim\epsilon^{\beta}$) and the correlation volume
($\xi_v\sim\epsilon^{-\bar{\nu}}$) were obtained as $\beta=1/2$
and $\bar{\nu}=2$, respectively~\cite{Kuramoto1984PTPS,
Daido1987JPA}, where $\epsilon$ is the reduced control parameter
and natural frequencies were randomly assigned from the Gaussian
distribution.
However, based on the FSS theory and heuristic arguments, the FSS
exponent $\bar{\nu}$ has been re-obtained as
$\bar{\nu}=5/2$~\cite{Hong2004n2005PRE+2007PRL}. It was taken into
account for size-dependent sample-to-sample fluctuations in
natural frequencies, but numerical confirmation was not entirely
satisfactory due to finite-size effects. Meanwhile, it has been
also reported that {\em thermal noise}, {\em quenched disorder} of
natural frequencies, and {\em link disorder} of coupled
oscillators, can also be relevant to the value of the FSS
exponent~\cite{Son2010PRE,Tang2011JSM,Hong2013PRE}.

In the absence of exact solutions, numerical tests are inevitable,
which is limited to finite systems related to computing
facilities. This issue has long been recognized in phase
transitions and critical phenomena. While FSS has played a crucial
role in its remedy, it requires the steady-steady limit of finite
systems, which takes quite a long computation time in the
numerical sense. Up to now, the FSS analysis of phase
synchronization has been carried out based on the steady-state
limiting data only. So one can naturally pose the following
question: What if there are only temporal data available? Is there
any systematic approach to deal with them? The answers will be
carefully addressed in this paper.

We propose an extended FSS form of the phase order parameter,
which provides another comprehensive view of synchronization with
the connection of dynamic scaling to FSS near and at the
criticality. In particular, we focus on how the order parameter
behaves in the true scaling regime before it gets into the steady
state, involved with the FSS exponent. Owing to the dynamic
scaling analysis, we successfully confirm the theoretical value
$\bar{\nu}=5/2$. Moreover, we also show $\bar{\nu}=2$, which is
clearly distinct from it in the presence of thermal noise. As a
final remark, we discuss the oscillatory behavior of the order
parameter in time with two scaling regimes. This occurs when the
KM starts at an incoherent state with fluctuation-free natural
frequencies by the regular sampling from the Gaussian
distribution.

It is well known that dynamic scaling is useful in nonequilibrium
systems such as surface growths~\cite{Barabasi1995}, cluster
aggregation models~\cite{Vicsek1984PRL}, and absorbing phase
transitions~\cite{Marro1999}. However, the dynamic scaling
analysis in synchronization models has not yet been studied
seriously to our knowledge.

The main purpose of this paper is to present dynamic scaling in
synchronization and to clarify its universality issue as
approaching the critical coupling strength.

This paper is organized as follows: In Sec.~\ref{model}, we
briefly review the ordinary KM and the conventional FSS theory of
the phase order parameter. In Sec.~\ref{DS}, we present the
dynamic scaling concept using the extended FSS theory and test it
with two completely different initial setups. The validity and the
universality issue of dynamic scaling are discussed in
Sec.~\ref{validity+universality} with numerical tests of thermal
noise and quenched disorder fluctuation. Finally, we conclude in
Sec.~\ref{summary} with a summary of our findings.

\begin{figure*}[]
\centering
\includegraphics[width=0.3\textwidth]{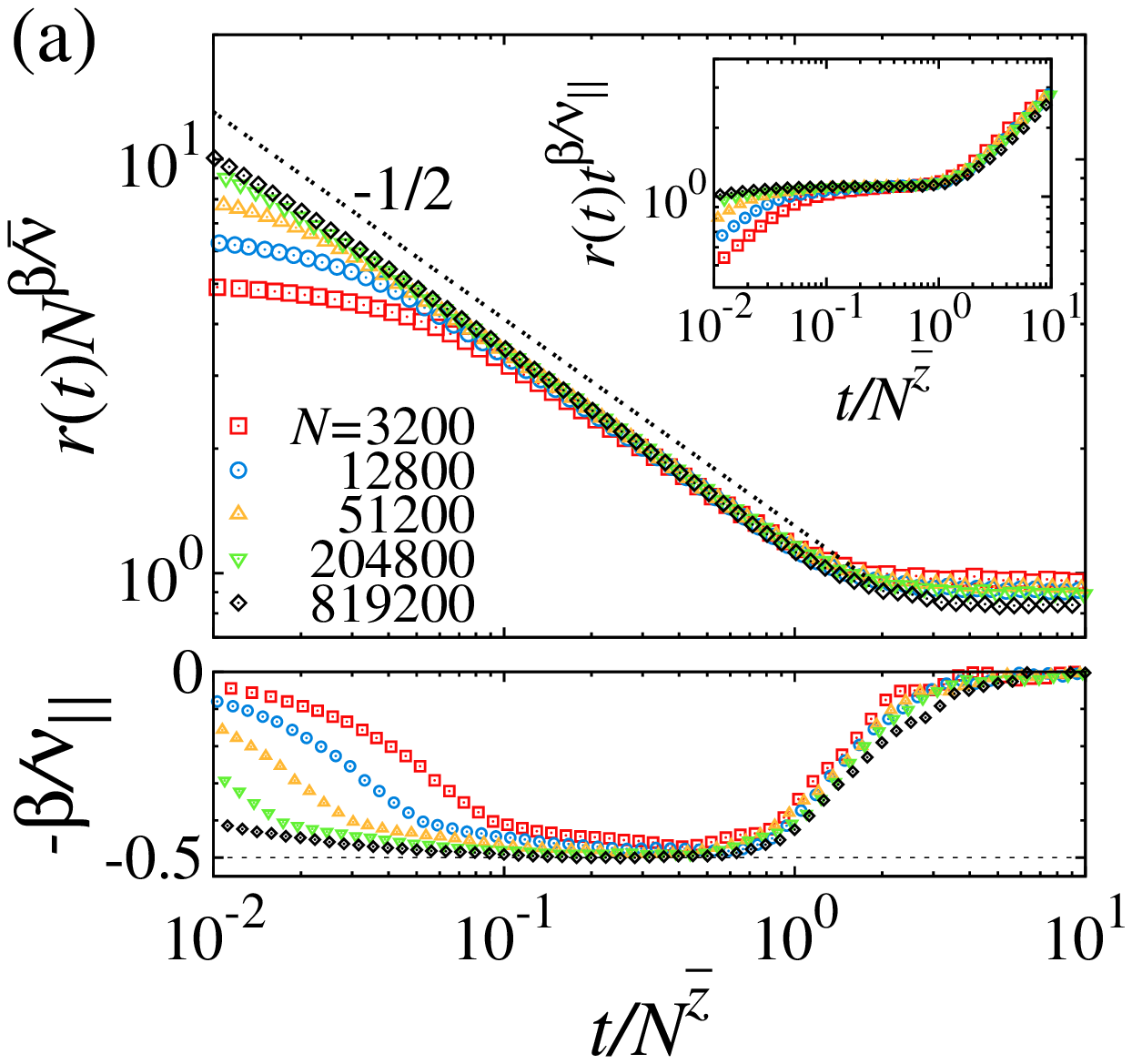}
\includegraphics[width=0.3\textwidth]{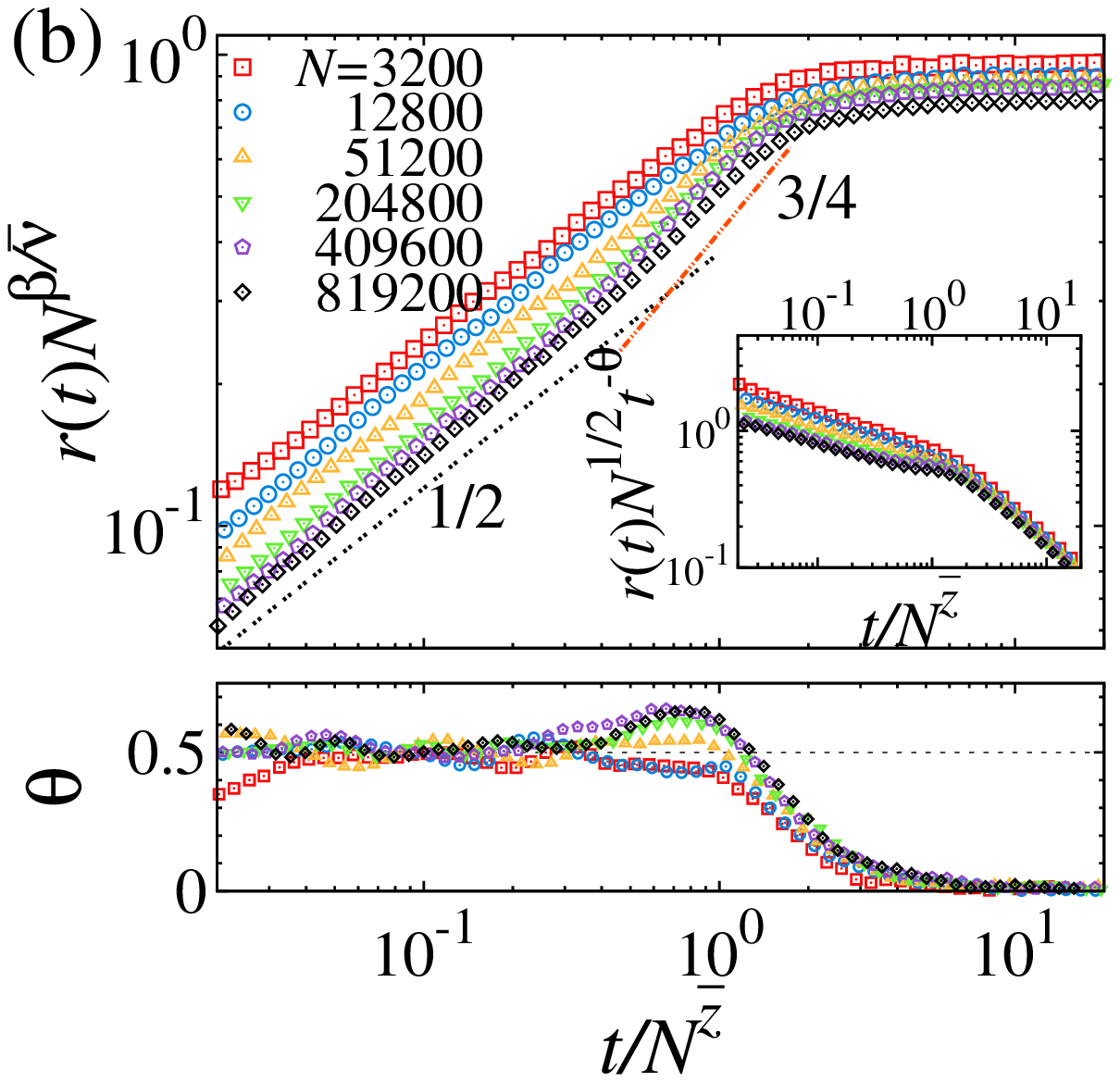}
\includegraphics[width=0.3\textwidth]{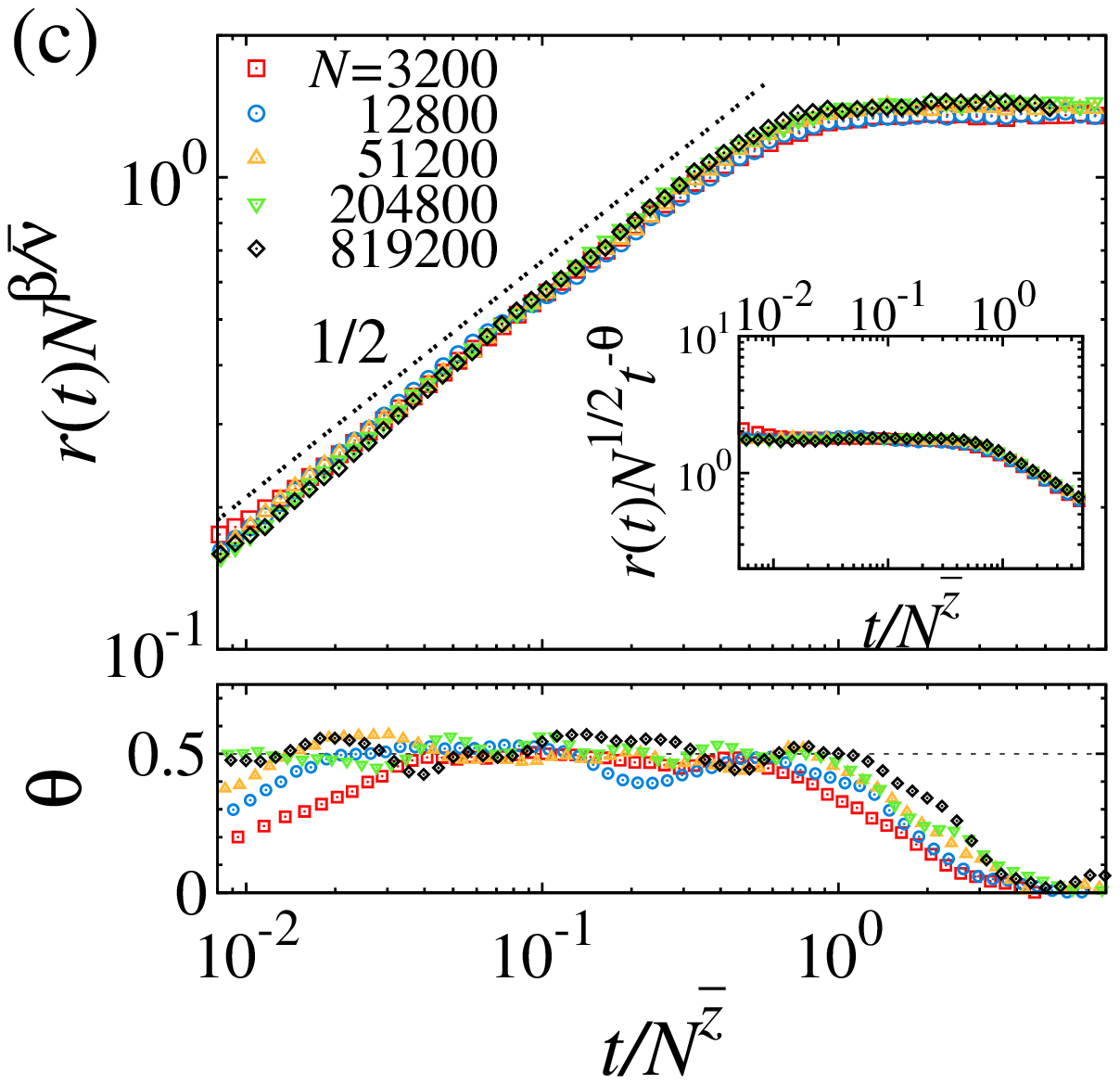}
\caption{(Color online) Scaling collapse of
Fig.~\ref{fig:DS4Gaussian-}: In the upper panel, ${\mathcal
{F}}(x)=r(t)N^{\alpha}$ is tested as main plots and
$f(x)=r(t)N^{\zeta}t^{y}$ as inset plots, where $x\equiv
t/N^{\bar{z}}$, $\alpha=\beta/\bar{\nu}$, and
$(\zeta,y)=[(0,~\beta/\nu_{||})~{\rm {for~
(a)}};~(1/2,-\theta)~{\rm {~for~(b)~and~(c)}}$] at $K=K_{\rm
c}(T)$ for various $N$. In the lower panel, the corresponding
effective exponents are also plotted using dynamic scaling with
the exponent set of
($\beta/\bar{\nu},~\bar{z},~\beta/\nu_{\parallel}\mbox{~or~}\theta$):
When the system starts (a) at a coherent [$r(0)= 1$] with
(1/5,~2/5,~1/2); (b) at an incoherent state [$r(0)\sim N^{-1/2}$]
with (1/5,~2/5,~3/4); (c) at the same state as (b) but containing
thermal noise ($T=0.1$) with (1/4,~1/2,~1/2).}
\label{fig:DS4Gaussian}
\end{figure*}
\begin{figure}[]
\centering
\includegraphics[width=0.48\columnwidth]{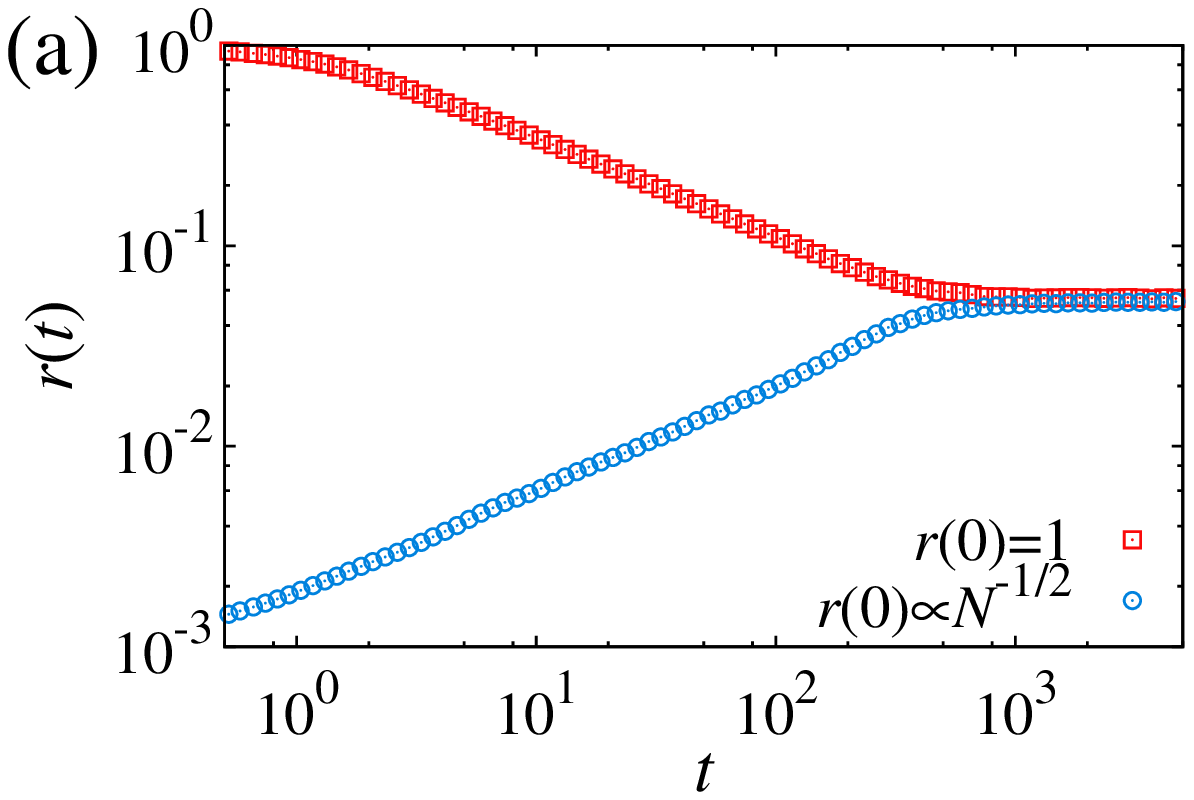}
\includegraphics[width=0.48\columnwidth]{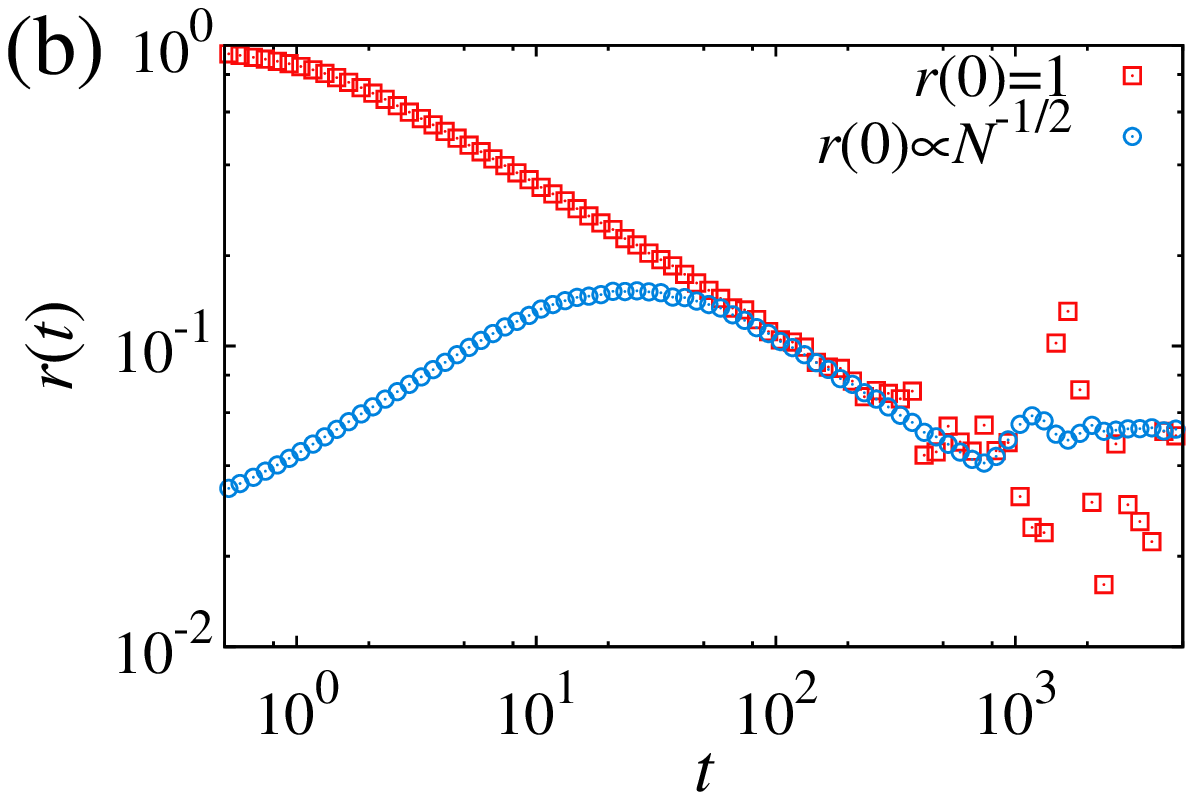}
\caption{(Color online) The noiseless ($T=0$) case: (a) for the
random sampling of $\{\omega_j\}$ at $N=102400$ and (b) for the
regular sampling of $\{\omega_j\}$ at $N=1600$ (relatively small
to the random case due to slow relaxation). Here circle (blue)
symbols start from an incoherent state and square (red) ones from
a coherent state.} \label{fig:test}
\end{figure}

\section{Model}
\label{model}

We begin with the KM~\cite{Kuramoto}, a paradigm of random
intrinsic frequency oscillators with the all-to-all coupling,
which is defined by the set of dynamic equations as
\begin{equation}
\frac{{\rm d}{\phi}_j(t)}{{\rm d}{t}} = \omega_j +
\frac{K}{N}\sum_{k=1}^{N}{\sin\big(\phi_k(t)-\phi_j(t)\big)},
\label{eq:KM-all-to-all}
\end{equation}
where $\phi_j(t)$ is the phase of the $j$-th oscillator at time
$t$ ($j,k=1,...,N$ for total number of oscillators), $\omega_j$ is
its time-independent natural frequency that follows the
distribution $g(\omega)$, and $K$ is the coupling strength. To
observe a second-order (continuous) synchronization transition, we
set $g(\omega)$ to be a Gaussian with zero mean and unit variance:
$g(\omega)=\frac{1}{\sqrt{2\pi}}\exp(-\frac{\omega^2}{2})$. It is
well-known that $\{\omega_j\}$ in the KM plays a role as quenched
disorder and its functional shape, $g(\omega)$, is relevant to the
nature of the synchronization transition~\cite{Pazo2005PRE}.
As $K$ increases, phase synchronization occurs at the critical
coupling strength $K_{\mathrm{c}}=\frac{2}{\pi
g(0)}(=\sqrt{8/\pi})$ \cite{Kuramoto}, which can be quantified by
a global complex-valued order parameter:
\begin{equation}
r(t)e^{i \psi(t)} \equiv \frac{1}{N} \sum_{k=1}^{N}e^{i
\phi_k(t)}. \label{eq:r}
\end{equation}

For the conventional FSS analysis, one collects the order
parameter $r$ only after it gets saturated to the steady-state
limiting value, where the time-averaged value is also taken,
denoted as $\langle r \rangle$, and the sample-averaged value over
the different sets of $\{\phi_j(0)\}$ at $t=0$ and $\{\omega_j\}$
is denoted as $[\langle r \rangle]$. To discuss dynamic scaling in
synchronization, we also focus on $r(t)$ (actually $[r(t)]$ used
to reduce statistical errors) for the whole regimes from the
dynamic state up to the steady state
[see~Fig.~\ref{fig:off-scaling} and Fig.~\ref{fig:DS4Gaussian-}].
It is already known that $r(t)$ grows exponentially far from the
criticality: $r(t)\sim\exp(at)$ before it saturates to $r_{\rm
sat}$ for $K \gg K_{\mathrm{c}}$~\cite{Strogatz1991JSP}. For $K
\ll K_{\mathrm{c}}$, it does not grow enough but fluctuates near 0
as much as $O(N^{-1/2})$. Moreover, the relaxation and decay
mechanism below $K_{\mathrm{c}}$ had been discussed with the
similarity of Landau damping~\cite{Strogatz1992PRL}. So the
naturally posed question is how it evolves near and at
$K=K_{\mathrm{c}}$.

In this paper, we trace the formation of synchronized clusters and
the cooperative behavior with time in the vicinity of $K_{\rm
c}~(\epsilon\equiv\frac{K-K_{\rm c}}{K_{\rm c}}=0)$, as the
correlation volume $\xi_v$ and the correlation time $\tau$ become
very large, compared to the subcritical regime $(\epsilon<0)$ and
the supercritical regime $(\epsilon>0)$, which algebraically decay
as $\xi_v \sim |\epsilon|^{-\bar{\nu}}$ and $\tau \sim
|\epsilon|^{-\nu_{\parallel}}$, respectively. However, $\xi_v\to
N$ in finite systems at $\epsilon=0$. As a result, $\tau\sim
N^{\bar{z}}$ with $\bar{z}={\nu_{\parallel}}/{\bar{\nu}}$.
Therefore, we are able to estimate the FSS exponent $\bar{\nu}$
using both temporal and static properties of the order parameter
from either $\bar{z}$ of the saturation time ($t_{\rm sat}\sim\tau
\sim N^{\bar{z}})$ or $\alpha \equiv \beta/\bar{\nu}$ of the
saturation value ($r_{\rm sat}\sim N^{-\alpha}$) as well as the
critical threshold $K_{\mathrm{c}}$ in two independent ways.

All numerical data presented here are obtained using the 4th order
Runge-Kutta method and $\mathrm{d}t$=0.01, which are averaged over
at least 500 samples, except Fig.~\ref{fig:off-scaling} in which
200 ensemble is enough.

\begin{figure*}[]
\centering
\includegraphics[width=0.3\textwidth]{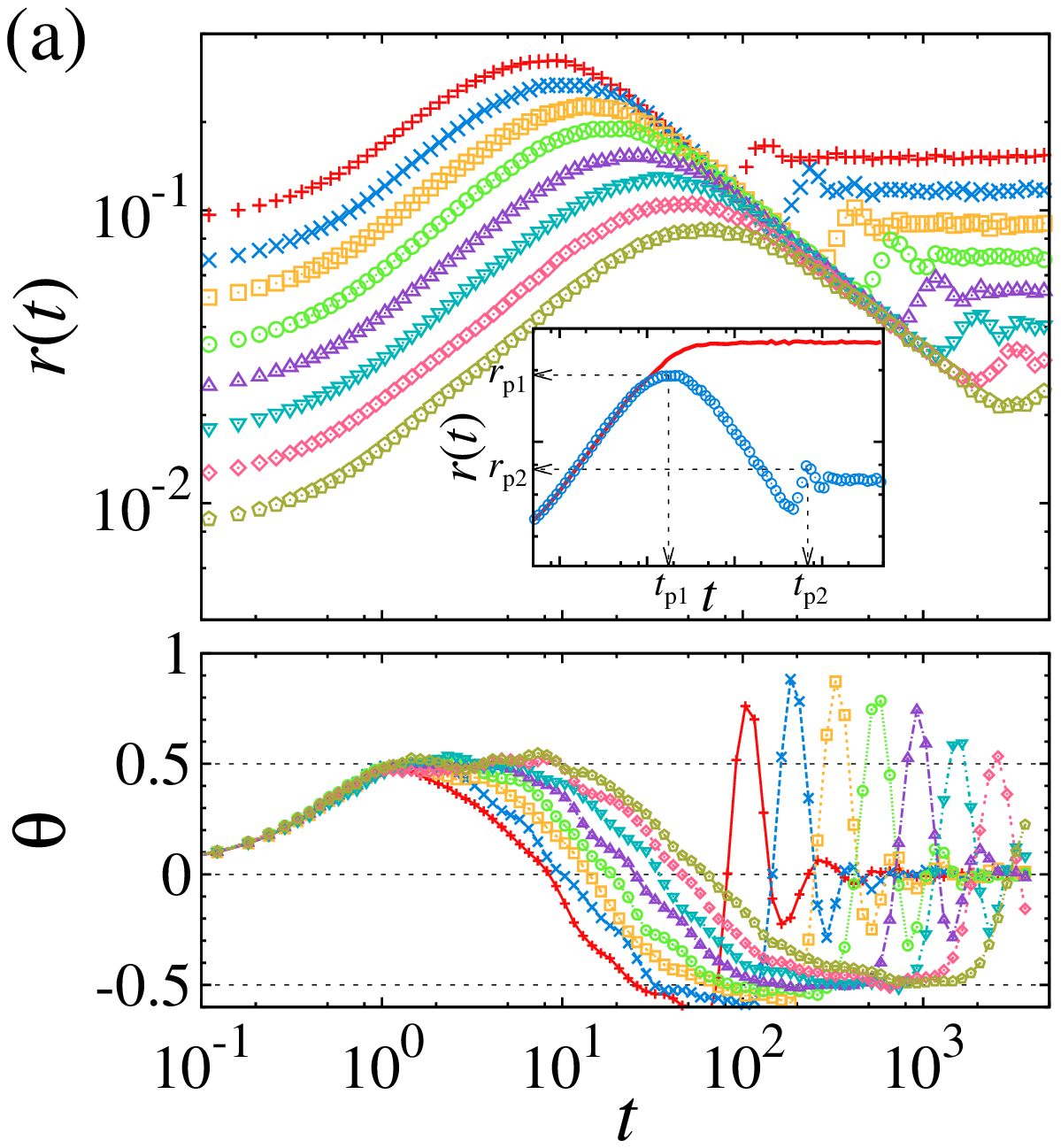}
\includegraphics[width=0.3\textwidth]{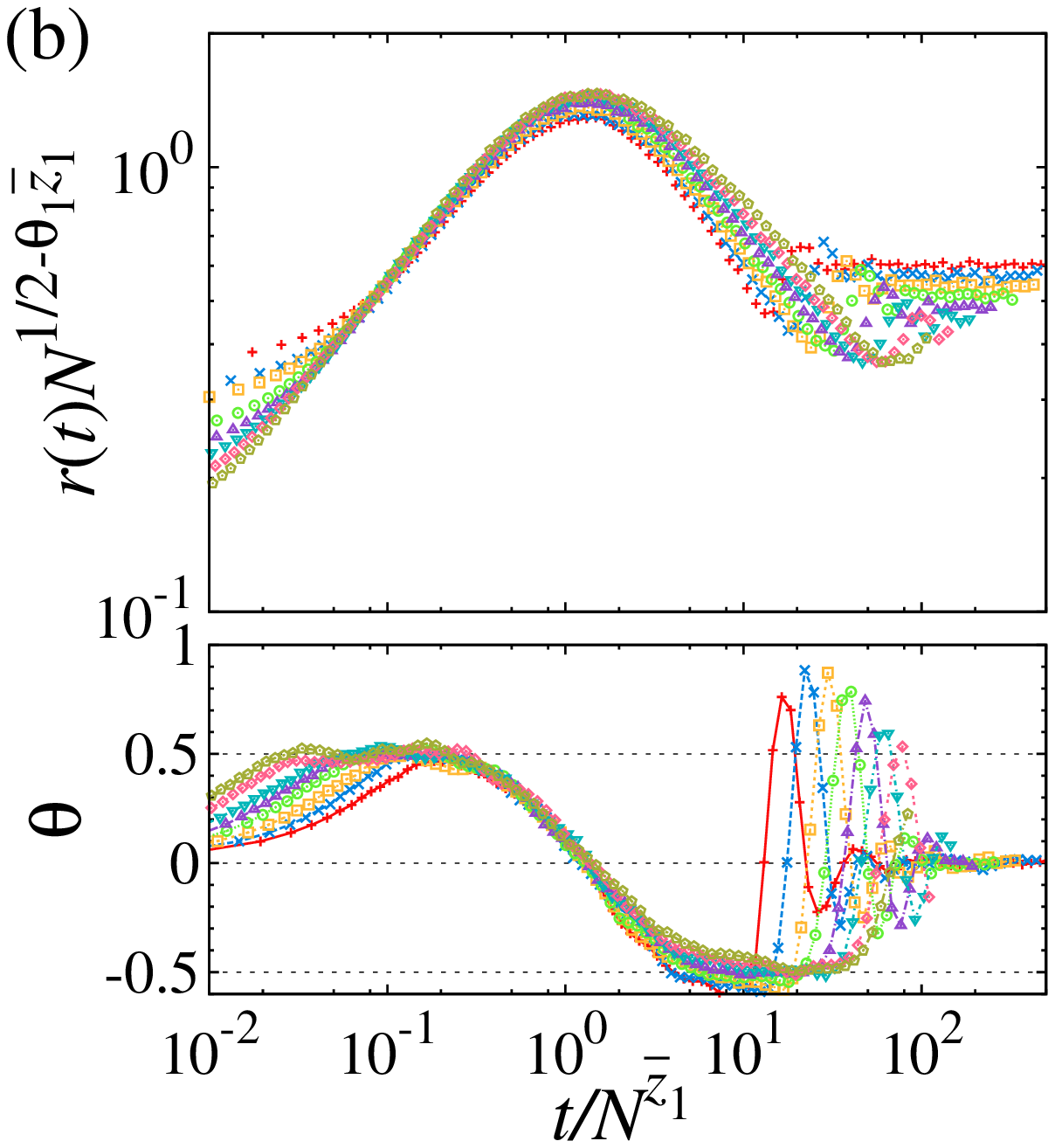}
\includegraphics[width=0.3\textwidth]{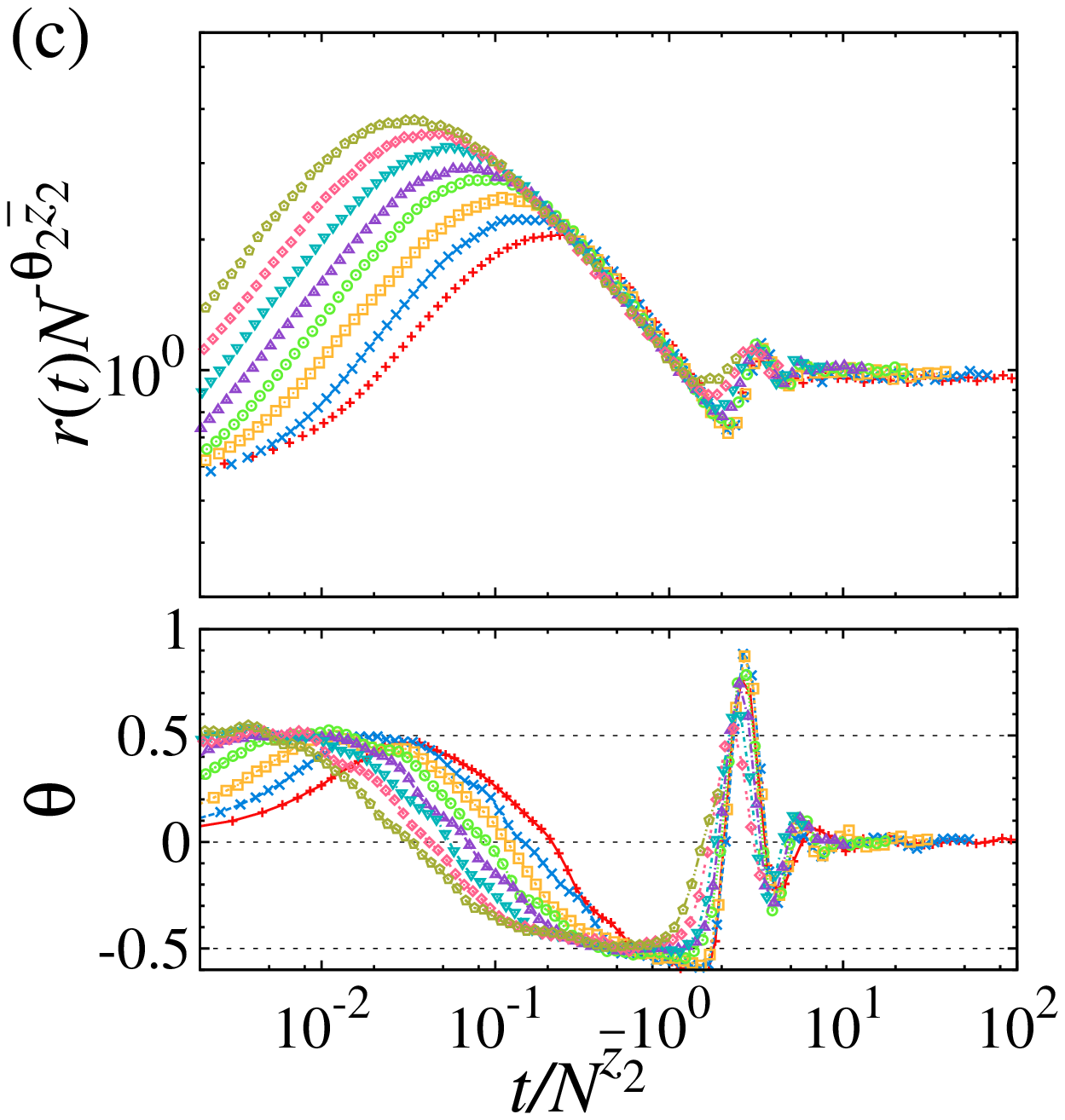}
\caption{(Color online) For the noiseless regular sampling of
$\{\omega_j\}$, dynamic scaling is tested at various $N$. (a)
Temporal behaviors of $r(t)$ (upper panel) and the corresponding
effective exponent (lower panel) are plotted, where
$N=100,~200,~\cdots,~12800$ from top to bottom. In the inset, the
random case (red, line) is compared with the regular one (blue,
symbol) with $N=800$. Two sets of data collapse of $r(t)$ are
shown for two different scaling regimes as well as those of
effective exponents (b) near the first peak with $\theta_1=1/2$
and $\bar{z}_1=2/5$, and (c) near and after the second peak with
$\theta_2=-1/2$ and $\bar{z}_2=4/5$.} \label{fig:regular-Gaussian}
\end{figure*}

\section{Dynamic Scaling}
\label{DS}

When a system exhibits self-similar dynamics at the criticality,
one can focus on dynamic scaling with a proper initial setup.

We revisit phase synchronization in the ordinary KM since the
values of $K_{\rm c}$ and $\beta$ are exactly known. Owing to that
fact, we easily test various properties and confirm the existence
of dynamic scaling. However, we note that the dynamic scaling
analysis is also powerful to indicate the location of $K_{\rm c}$
[see Fig.~\ref{fig:off-scaling}]. Furthermore, we discuss the
universality issue in synchronization, related to the relevance of
thermal and link-disorder fluctuations of oscillators against two
different sampling methods of natural frequencies.

Two different initial conditions of the KM are chosen to start
either at a fully coherent state [where $\phi_j(0)=\phi_o$: an
arbitrary angle, independent of $j$, so $r(0)=1$] or at an
incoherent state [where $\phi_j(0)\in [0,2\pi)$ is random, so
$r(0)\sim N^{-1/2}$]. For a given value of $K$, $r(t)$ evolves
either exponentially or algebraically up to $\tau\equiv t_{\rm
sat}$, which is also subject to the system size $N$.

Based on the FSS theory and thermodynamic limiting results as
$N\to \infty$: $t_{\rm sat}\sim \epsilon^{-\nu_{\parallel}}$ with
$\nu_{\parallel}=1$ and $r_{\rm sat}\sim \epsilon^{\beta}$ with
$\beta=1/2$, which is also numerically confirmed in
Fig.~\ref{fig:off-scaling}. So the extended FSS to dynamic scaling
can be rewritten near and at $\epsilon=0$ as
\begin{equation}
r(t,N,\epsilon)=b^{-\alpha}r_b(b^{-\bar{z}}t,b^{-1}N,
b^{1/\bar{\nu}}\epsilon), \label{eq:extended-FSS}
\end{equation}
where $b$ is an arbitrary scaling factor and
$\alpha\equiv\beta/\bar{\nu}$. In the steady-state limit
($t\to\infty$), Eq.~(\ref{eq:extended-FSS}) is exactly the same as
the earlier FSS form, $r(\epsilon, N) = N^{-\alpha}f(\epsilon
N^{1/\bar{\nu}})$~\cite{Hong2004n2005PRE+2007PRL}.

Equation~(\ref{eq:extended-FSS}) can also be rewritten as the
dynamic scaling form with two variables, $t$ and $N$, as
$N\to\infty$ ($b=t^{1/\bar{z}}$) or as $t\to\infty$ ($b=N$):
\begin{equation}
r(t,N)=t^{-\alpha/\bar{z}}f(t/N^{\bar{z}})=N^{-\alpha}\mathcal{F}(t/N^{\bar{z}}),
\label{eq:DS}
\end{equation}
which is numerically confirmed (see
Figs.~\ref{fig:DS4Gaussian-}-\ref{fig:DS4Gaussian}). Here
$\alpha=\beta/\bar{\nu}$ from $r_{\rm sat}$ and
$\bar{z}=\nu_{\parallel}/\bar{\nu}$ from $t_{\rm sat}$.

To confirm that the transition is continuous and discuss how the
initial setup affects dynamic scaling at the transition in detail,
two completely different configurations are considered, which
correspond to Fig.~\ref{fig:DS4Gaussian}~(a) and (b) for the
ordinary KM starting from a fully coherent state and from a random
(incoherent) state, respectively.

The below form of dynamic scaling describes that the KM initially
starts at $r(0)=1$. As time elapses, the order parameter decays as
a power law, denoted as $r_{\downarrow}(t,N)$:
\begin{eqnarray}
r_{\downarrow}(t,N)
&=&t^{-\alpha/\bar{z}}f_{\downarrow}(t/N^{\bar{z}})
=N^{-\alpha}\mathcal{F}_{\downarrow}(t/N^{\bar{z}})\nonumber\\
&\sim &
\begin{cases}
t^{-\alpha/\bar{z}} & \text{for } t_{\times}<t\ll t_{\rm sat}(\sim N^{\bar{z}}),\\
N^{-\alpha} & \text{for } t\gg t_{\rm sat},\\
\end{cases}
\label{eq:DS-synch}
\end{eqnarray}
where $f_{\downarrow}(x)$ is constant for $x\ll 1$ in the true
scaling regime ($t_{\times}<t\ll t_{\rm sat}$) after the transient
regime ($t<t_{\times}$ when the initial condition effect exists;
$t_{\times}$ is independent of $N$ in general), and
$f_{\downarrow}(x)\sim x^{\alpha/\bar{z}}$ for $x\gg 1$ in the
saturation regime ($t\gg t_{\rm sat}\sim N^{\bar{z}}$; when the
system-size dependence only exists) [see
Fig.~\ref{fig:DS4Gaussian-} (a) and Fig.~\ref{fig:DS4Gaussian}
(a)].

If one chooses an initial configuration starting at an incoherent
state with $N$-dependent randomness [$r(0)\sim N^{-1/2}$], the
order parameter increases in a trivial power law to wash out such
randomness after the transient regime, and then it exhibits true
scaling. Therefore, Eq.~(\ref{eq:DS-synch}) should be modified due
to $N$-dependent trivial offset ($\sim N^{-1/2}$) and trivial
temporal scaling ($\sim t^{1/2}$), denoted as $r_{\uparrow}$ for
convenience, as follows:
\begin{eqnarray}
r_{\uparrow}(t,N)&=&
N^{-1/2}t^{\theta}f_{\uparrow}(t/N^{\bar{z}})=N^{-\alpha}\mathcal{F}_{\uparrow}(t/N^{\bar{z}})\nonumber\\
& \sim &
\begin{cases}
N^{-1/2}t^{1/2} & \text{for } t_{\times}< t<t_{\rm cross},\\
N^{-1/2}t^{\theta} & \text{for } t_{\rm cross}\ll t\ll t_{\rm sat},\\
N^{-\alpha} & \text{for } t\gg t_{\rm sat},\\
\end{cases}
\label{eq:DS-desynch}
\end{eqnarray}
where $f_{\uparrow}(x)$ is constant for $x_*(\equiv t_{\rm
cross}/N^{\bar{z}})\ll x\ll 1$ in the true scaling regime, and
$f_{\uparrow}(x)\sim x^{({\alpha-\frac{1}{2}})/\bar{z}}$ for $x\gg
1$ in the saturation regime [see
Fig.~\ref{fig:DS4Gaussian-}~(b),(c) and
Fig.~\ref{fig:DS4Gaussian}~(b),(c)].

Figure~\ref{fig:DS4Gaussian-}~(b) [see
Fig.~\ref{fig:DS4Gaussian}~(b) as well] shows very long transient
trivial scaling in the time evolution of $r(t)$ due to random
phases at $t=0$, $r(t)\sim N^{-1/2}t^{1/2}$. This lasts up to
$t_{\rm cross}$ until the random initial condition effect is
washed out and the system exhibits true scaling with
$N^{-1/2}t^{\theta}$. In order to resolve this universality issue,
one needs to find the crossover time $t_{\mathrm{cross}}$
accurately as well as the true scaling behavior. It is definitely
not an easy task and sometimes extremely tricky if the window of
two consecutive scaling regimes is narrow because one scaling
interferes with the other one.

From the fact that at a continuous transition the steady state
should be the same, irrespective of initial setups [see
Fig.~\ref{fig:test}~(a)], we derive a scaling relation among
$\alpha(=\beta/\bar{\nu}),~\theta,$ and
$\bar{z}(=\nu_{\parallel}/\bar{\nu})$ as
$\frac{1}{2}-\theta\bar{z}=\alpha$ in $r_{\rm sat}\sim
N^{-\alpha},~r_{\uparrow}(t)N^{1/2}\sim t^{\theta},$ and $t_{\rm
sat}\sim N^{\bar{z}}$, respectively. This is equivalent to
$\theta=(\frac{1}{2}-\alpha)/\bar{z}=(\frac{\bar{\nu}}{2}-\beta)/\nu_{\parallel}$.
Hence, $r_{\uparrow}(t)$ for the random sampling of $\{\omega_j\}$
with the random choice of $\{\phi_j(0)\}\in [0,2\pi)$ is
characterized by two different length scales, unlike the
conventional temporal behavior in a simple power-law manner. It is
because it is involved with two different dynamic exponents, which
is attributed to the finite-size effect and the crossover from
$t^{1/2}$ to $t^{3/4}$ at $t_{\rm cross}$ as time elapses.

The true dynamic exponent $\bar{z}$ related to the true FSS
exponent $\bar{\nu}$ in the long-time regime after the crossover
yields $\tau\sim N^{\bar{z}}$ where $\bar{z}=1/\bar{\nu}=2/5$ with
$\nu_{\parallel}=1$ in networks, only observed in sufficiently
large system sizes. Otherwise, the crossover scaling of
$\bar{z}=1/\bar{\nu}=1/2$ is only detected, which is related to
thermal noise [see Fig.~\ref{fig:comparison}]. This anomalous
dynamic scaling of $r_{\uparrow}(t)$ is resolved with thermal
noise $\eta_j(t)$ using the modified KM~\cite{Son2010PRE}:
\begin{equation}
 \frac{{\rm d}{\phi}_j(t)}{{\rm d}{t}} = \omega_j +
\frac{K}{N}\sum_{k=1}^{N}{\sin\big(\phi_k(t)-\phi_j(t)\big)} +
\eta_j(t),
\label{KM+T}
\end{equation}
 where $\langle\eta_j(t)\rangle=0$ and
$\langle\eta_j(t)\eta_k(t')\rangle=2T\delta_{jk}\delta(t-t')$. In
the modified KM, we observe that the conventional dynamic scaling
governed by random fluctuations with $\bar{z}=1/\bar{\nu}=1/2$ as
expected [see Fig.~\ref{fig:DS4Gaussian-}~(c) and
Fig.~\ref{fig:DS4Gaussian}~(c)].

Using the KM with various settings, we discuss the universality of
the dynamic exponent in true scaling.

\section{Effects of Noise and Disorder}
\label{validity+universality}

In order to discuss the validity of our conjecture on the dynamic
scaling form, it is necessary to test the relevance of thermal
noise and the type of disorder in the KM as discussed in the FSS
theory~\cite{Son2010PRE, Tang2011JSM, HP+Hong}. In the presence of
thermal noise, it is always relevant, irrespective of disorder
type. So it changes the value of $\bar{\nu}=1/\bar{z}$ with
$\nu_{\parallel}=1$ from $\bar{\nu}=5/2$ to $\bar{\nu}=2$ (see
Table~\ref{table:FSS}).

Compared to the case of the noiseless ($T=0$) random sampling [see
Fig.~\ref{fig:DS4Gaussian-}~(b) and
Fig.~\ref{fig:DS4Gaussian}~(b)], $r_{\uparrow}(t)$ for the noisy
case exhibits clean dynamic scaling [see
Fig.~\ref{fig:DS4Gaussian-}~(c) and
Fig.~\ref{fig:DS4Gaussian}~(c)] with $\bar{\nu}=2$. This
distinction of these two cases plays a key role in detecting the
true scaling regime ($t \gg t_{\rm cross}$) for the case of
noiseless random sampling [see Fig.~\ref{fig:comparison}].
However, the window of the true scaling regime is somehow quite
short (at most one decade) and hardly observable in smaller
systems, implying that the case of noiseless random sampling is
hardly distinguishable with the noisy one in numerical senses
unless $N$ is big enough. This is why some numerical results
reported $\bar{\nu}=2$ (not $\bar{\nu}=5/2$) even for the
noiseless case.

Based on our extensive numerical simulation results,
$r_\uparrow(t)$ in bigger systems at least $N\ge 204,800$ exhibit
their own true scaling regime clearly [see
Fig.~\ref{fig:DS4Gaussian}(b),(e), and Fig.~\ref{fig:comparison}].
This is why one cannot observe true scaling in smaller systems
($N<N_{\mathrm{cross}}$), which $t_{\rm cross}(N)\ge t_{\rm
sat}(N)$ due to finite-size corrections to scaling. Note that
$N_{\mathrm{cross}}=O(10^5)$ can be estimated from
$r_{\mathrm{sat}}(\epsilon, N)=N^{-1/5} f(\epsilon N^{2/5})$ and
$r_{\mathrm{sat}}\ll 1$ at $\epsilon=0$.

To discuss the relevance of natural frequency sampling (quenched
disorder type) in dynamic scaling as well as the initial setups,
we revisit the KM in the absence of thermal noise. If
$\{\omega_j\}$ is regularly generated by $\omega_j = \sqrt{2}
\mathrm{erf}^{-1} \left( -1+\frac{2j-1}{N} \right)$, it plays a
role as ``sample-to-sample fluctuation-free" quenched disorder in
the system.

\begin{figure}[t]
\centering
\includegraphics[width=0.625\columnwidth]{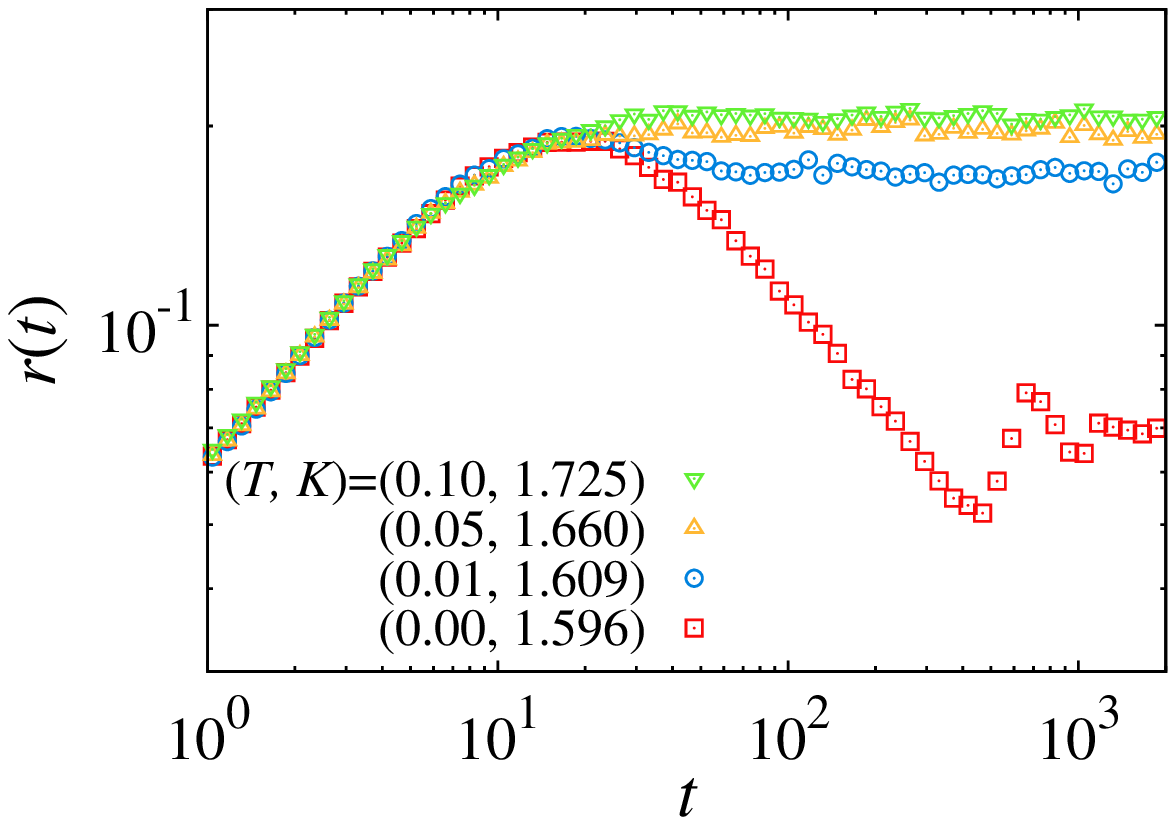}\\
\includegraphics[width=0.625\columnwidth]{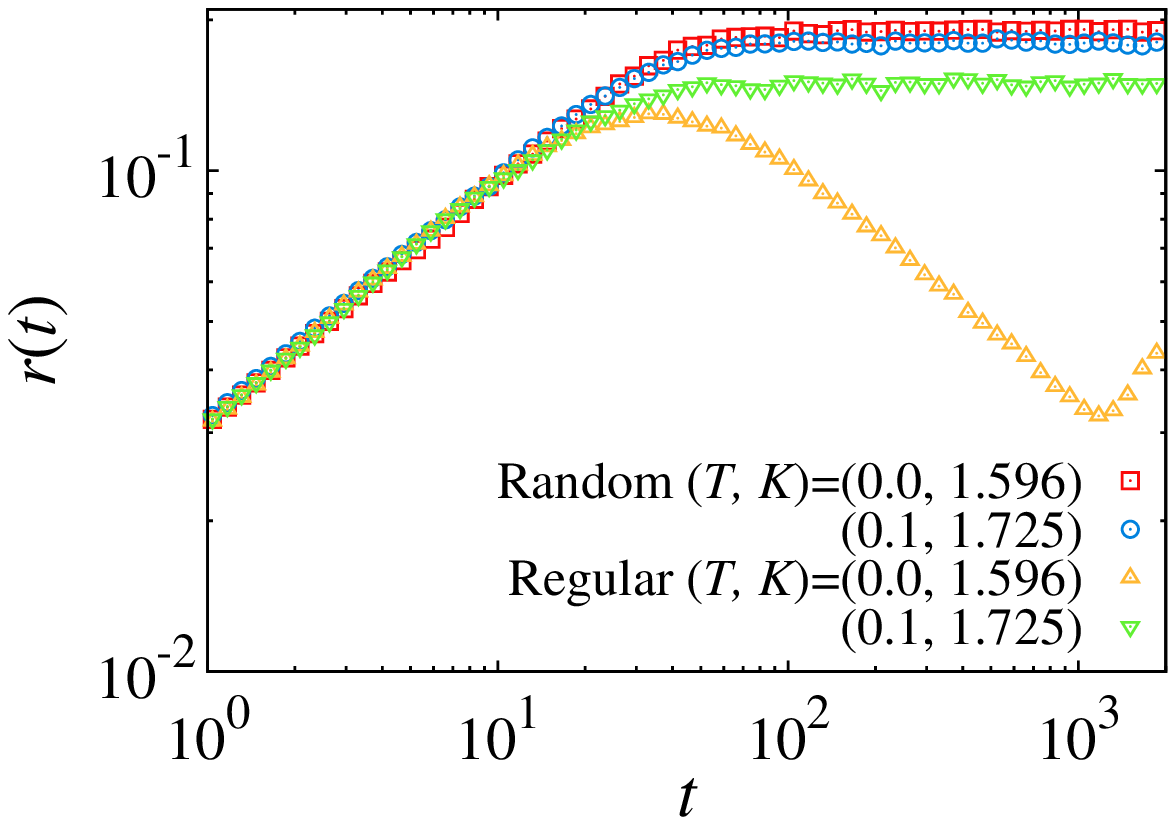}
\caption{(Color online) Critical behaviors of $r_{\uparrow}(t)$ at
$K=K_{\rm c}(T)$: at $N=800$ for the regular sampling of
$\{\omega_j\}$ for the upper panel and at $N=3200$ for four
different cases that are described in Table~\ref{table:FSS} with
1000 ensembles for the lower panel. However, the steady-state
limiting value in finite systems, $r_{\rm sat}(N)$, seems to
depend on the value of $T$ for the regular sampling case of
$\{\omega_j\}$ if the effect of $T$ is relatively small, compared
to the effect of $N$.} \label{fig:validity}
\end{figure}

For this regular sampling [see Fig.~\ref{fig:regular-Gaussian}],
$r_{\uparrow}(t)$ exhibits very interesting damped oscillation,
rather than anomalous crossover scaling for the random sampling
case. However, if a system exhibits a continuous phase transition,
the steady-state limit should be independent of initial setups.
Through Fig.~\ref{fig:test}, we confirm that the order parameter
for the noiseless ($T=0$) case starting two completely different
initial setups has the same value in the steady-state, and we find
that the anomalous oscillatory behavior exists for the regular
sampling of $\{\omega_j\}$ starting with an incoherent state. The
comparison with the noisy case ($T\ne 0$) is shown in
Fig.~\ref{fig:validity}. In the inset of
Fig.~\ref{fig:regular-Gaussian}~(a), the heights of two largest
peaks at the corresponding times are taken as indicators,
respectively.

\begin{figure}[]
\centering
\includegraphics[width=0.925\columnwidth]{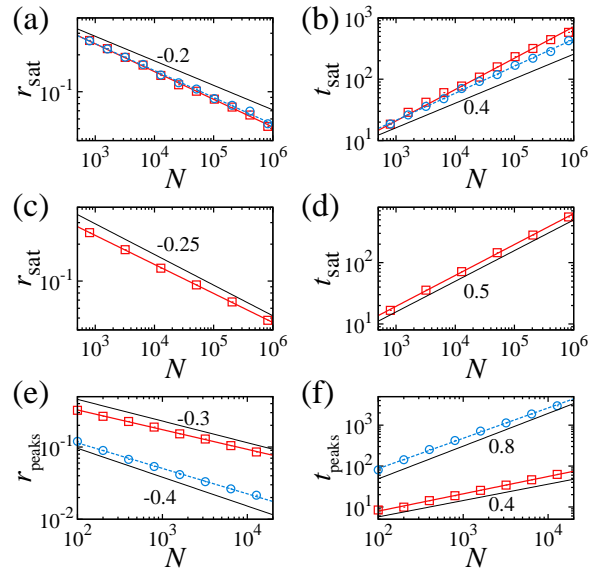}
\caption{(Color online) FSS data analysis for the saturated values
with $r_{\rm sat}\sim N^{-\alpha}$ and $t_{\rm sat}\sim
N^{\bar{z}}$: (a) and (b) for the noiseless random case, where
blue circle (red square) symbols represent the data starting at a
fully coherent (incoherent) state, and the slope sets
$(-\alpha,~\bar{z})$ of naive fitting lines correspond to (-0.22,
0.44) for circles and (-0.23, 0.51) for squares; (c) and (d) for
the noisy random case with $T=0.1$, where blue square symbols
represent the data starting at an incoherent state, and the slope
set of naive fitting lines corresponds to (-0.24, 0.50); (e) and
(f) for the noiseless regular case, where red square (blue circle)
symbols represent the scaling properties of the first (second)
peak in $r_{\uparrow}(t)$ [described in the inset of
Fig.~\ref{fig:regular-Gaussian}(a)], and the slope sets of naive
fitting lines correspond to (-0.27, 0.42) for squares and (-0.35,
0.74) for circles. } \label{fig:numerics}
\end{figure}

Based on numerical tests as shown in
Fig.~\ref{fig:regular-Gaussian}~(b),(c) and
Fig.~\ref{fig:numerics}~(e),(f), we find that
$(r_{\mathrm{p}1}\sim N^{-\alpha_1},~ t_{\mathrm{p}1}\sim
N^{\bar{z}_1})$ at the first largest one and $(r_{\mathrm{p}2}\sim
N^{-\alpha_2},~t_{\mathrm{p}2}\sim N^{\bar{z}_2})$ at the second
largest one, with ($\alpha_1=3/10$, $\bar{z}_1= 2/5$) with
$\theta_1=1/2$ for the first scaling regime and ($\alpha_2\simeq
2/5$, $\bar{z}_2\simeq 4/5$) with $\theta_2\simeq -1/2$ for the
second one. We conjecture the following scaling relations:
$\alpha_1=1/2-\theta_1\bar{z}_1$ and
$\alpha_2=-\theta_2\bar{z}_2$. As a result,
Eq.~(\ref{eq:DS-desynch}) should be modified to the following two
forms:
\begin{widetext}
\begin{eqnarray}
r_{\uparrow,\mathrm{p}1}(t,N) &=&
N^{-1/2}t^{\theta_1}f_{\uparrow,\mathrm{p}1}(t/N^{\bar{z}_1})=N^{-\alpha_1}\mathcal{F}_{\uparrow,\mathrm{p}1}(t/N^{\bar{z}_1})
\sim
\begin{cases}
N^{-1/2}t^{\theta_1} &\text{for } t_{\times}< t \ll t_{\mathrm{p}1}\sim N^{\bar{z}_1},\\
N^{-\alpha_1} & \text{at } t=t_{\mathrm{p}1}\sim N^{\bar{z}_1},\\
\end{cases}
\\
r_{\uparrow,\mathrm{p}2}(t,N) &=&
t^{\theta_2}f_{\uparrow,\mathrm{p}2}(t/N^{\bar{z}_2})=N^{-\alpha_2}\mathcal{F}_{\uparrow,\mathrm{p}2}(t/N^{\bar{z}_2})
\sim
\begin{cases}
t^{\theta_2} & \text{for } t_{\mathrm{p}1} \ll t \ll t_{\mathrm{p}2}\sim N^{\bar{z}_2},\\
N^{-\alpha_2} & \text{for } t\gg t_{\mathrm{p}2},
\end{cases}
\label{eq:DS-desynch-regular}
\end{eqnarray}
\end{widetext}
where $f_{\uparrow,\mathrm{p}1}(x)$ is constant for $x \ll 1$,
$f_{\uparrow,\mathrm{p}1}(x)\sim x^{-\theta_1}$ for $x \gg 1$ and
$f_{\uparrow,\mathrm{p}2}(x)$ is constant for $x \ll 1$,
$f_{\uparrow,\mathrm{p}2}(x)\sim x^{-\theta_2}$ for $x \gg 1$.
Figure~\ref{fig:regular-Gaussian}~(b) and (c) correspond to the
scaling function
${\mathcal{F}}_{\uparrow}(t/N^{\bar{z}})=r_{\uparrow}(t)N^{\alpha}$
in Table~\ref{table:FSS}.

Unlike the random sampling of $\{\omega_j\}$, the regular one has
not been fully understood except for the nontrivial value of the
FSS exponent ($\bar{\nu}\simeq 5/4$ reported
in~\cite{Son2010PRE,Tang2011JSM,Hong2013PRE}). Our dynamic scaling
results would give a hint to find the correct value of $\bar{\nu}$
but also address how and when the effect of initial condition is
washed out in $r(t)$ [see Fig.~\ref{fig:test}].

\begin{table*}[]
\caption{All critical exponents are summarized with the earlier
FSS results and our conjecture for dynamic scaling. For the
noiseless regular sampling of $\{\omega_j\}$, $\theta(t)=1/2$ for
$t \ll t_{\mathrm{p}1}(\sim N^{\bar{z}_1}$ with $\bar{z}_1\simeq
2/5$); -1/2 for $t_{\mathrm{p}1} \ll t \ll t_{\mathrm{p}2}(\sim
N^{\bar{z}_2}$ with $\bar{z}_2\simeq 4/5$) [see
Fig.~\ref{fig:regular-Gaussian} and Fig.~\ref{fig:numerics}]. For
all the cases, the scaling function ${\mathcal
F}_{\uparrow}(x)={\rm constant}$ for $x\gg 1$, where $x\equiv
t/N^{\bar{z}}$.}
\begin{tabular}{cccccc}
\hline\hline Sampling & ~~Noise~~ & Steady State & ~~Dynamic State
~&~Dynamic Scaling & Scaling Function\\
\\
$\{\omega_j\}$ from $g(\omega)$
&$T$&$(\beta/\bar{\nu},~1/\bar{\nu})$  &
$(\beta/\nu_{\parallel},~\theta,~\bar{z})$&starting at
$r(0)~N^{-1/2}$
&${\mathcal F}_{\uparrow}(x)$\\
\hline random & $T=0$ & (1/5, 2/5) & (1/2, 3/4, 2/5)
&$r_{\uparrow}(N,t)=N^{-2/5}{\mathcal F}_{\uparrow}(t/N^{2/5})$&$~x^{3/4} {\rm~for~}x_{\rm cross}\ll x\ll 1$\\
\\
&$T\ne 0$ & (1/4, 1/2) & (1/2, 1/2, 1/2)
&$r_{\uparrow}(N,t)=N^{-1/4}{\mathcal F}_{\uparrow}(t/N^{1/2})$&$~x^{1/2} {\rm~for~}x_{\times}\ll x\ll 1$\\
\\
regular & $T=0$ & $~~$(2/5, 4/5)\footstar{These values are from
Refs.~\cite{Hong2004n2005PRE+2007PRL,Son2010PRE,Tang2011JSM,Hong2013PRE}.}
& (1/2, 1/2, 2/5) & $r_{\uparrow, {\rm p1}}(N,t)=N^{-3/10}{\mathcal F}_{\uparrow,{\rm p1}}(t/N^{2/5})$&$~x^{1/2} {\rm~for~} x_{\times}\ll x\ll x_{\rm p1}$\\
&&& (1/2, -1/2, 4/5) & $r_{\uparrow,{\rm p2}}(N,t)=N^{-2/5}{\mathcal F}_{\uparrow,{\rm p2}}(t/N^{4/5})$&$~x^{-1/2} {\rm~for~} x_{\rm p1}\ll x\ll x_{\rm p2}$\\
\\
& $T\ne 0$ & (1/4, 1/2) & (1/2, 1/2, 1/2)
&$r_{\uparrow}(N,t)=N^{-1/4}{\mathcal F}_{\uparrow}(t/N^{1/2})$&$~x^{1/2} {\rm~for~}x_{\times}\ll x\ll 1$\\
\hline\hline
\end{tabular}
\label{table:FSS}
\end{table*}

Furthermore, the origin of oscillatory behaviors in dynamic
scaling is still under investigation. Figure~\ref{fig:validity}
shows that it is completely gone once thermal noise is turned on.
Most recently, it has been also reported in~\cite{Hong2013PRE}
that link fluctuations of oscillator networks generate effective
fluctuations of natural frequencies, which means the absence of
oscillatory behaviors once random fluctuations in links of
oscillator networks are considered. Such a change is also
numerically observed. A more detailed investigation for dynamic
scaling~\cite{unpublished} will be provided elsewhere to complete
the discussion of the universality issue in synchronization as
well as the transition nature against the distribution type of
natural frequencies.

Finally, we discuss how the strength of thermal noise ($T$)
affects dynamic scaling of $r_{\uparrow}(t)$ at $K=K_{\rm c}(T)$,
which is based on Fig.~\ref{fig:validity}. Once we turn on thermal
noise, the oscillatory behavior for the noiseless case is washed
out. For four different cases that are described in
Table~\ref{table:FSS}, we also compare one with another. Moreover,
all the FSS data analysis and dynamic scaling results are
summarized in Fig.~\ref{fig:numerics} and Table~\ref{table:FSS} in
detail manner as possible.

\section{Summary and Discussions}
\label{summary}

In conclusion, we have systematically explored dynamic scaling of
synchronization in the Kuramoto model, and investigated scaling
relations between our results and the earlier FSS ones. We also
found that dynamic scaling properties can also clearly locate the
critical coupling strength of synchronization and estimate the
values of critical exponents. As a final remark, we addressed how
the initial phases of oscillators and the generation method of
natural frequency sequences affect dynamic scaling and the FSS
exponent, which were numerically confirmed.

The merit of dynamic scaling, similar to the earlier work on the
short-time behavior of the two-dimensional $\phi^4$
theory~\cite{Zheng1999PRL}, is to provide another comprehensive
view of synchronization by the time evolution of the order
parameter before the system reaches the steady state against
various initial setups. This offers a guideline how to analyze a
phase synchronization transition in finite systems without any
steady-state limiting results.

We believe that dynamic scaling provides rich information in
analyzing real systems, including the transition nature and the
universality issue.

\section*{ACKNOWLEDGMENTS}

This work was supported by the NRF grant funded by the Korean
Government (MEST/MSIP) (No. 2011-0011550/2013-027911) (M.H.); (No.
2010-0015066) (C.C., B.K.). M.H. would also acknowledge the
generous hospitality of KIAS, where fruitful discussion with H.
Park, H. Hong, J. Um, and S. Gupta could be had, and its support
through the Associate Member Program by the MEST.


\begin{thebibliography}{99}

\bibitem{old-intro+review}
P. Barbara, A.B. Cawthorne, S.V. Shitov, and C.J. Lobb, Phys. Rev.
Lett. {\bf 82}, 1963~(1999);
I.Z. Kiss, Y.M. Zhai, and J.L. Hudson, Science {\bf 296},
1676~(2005);
J.A. Acebr\'{o}n {\it et. al.}, Rev. Mod. Phys. {\bf 77},
137~(2005).

\bibitem{Rohden2012PRL+Doefler2012PNAS+Motter2013NPHYS}
M. Rohden, A. Sorge, M. Timme, and D. Witthaut, Phys. Rev. Lett.
{\bf 109}, 064101 (2012); F. D\"{o}rfler, M. Chertkov, and F.
Bulllo, PNAS {\bf 110}, 2005 (2012); A.E. Motter, S.A. Myers, M.
Anghel, and T. Nishikawa, Nat. Phys. {\bf 9}, 191 (2013).

\bibitem{Martens2013PNAS}
E.A. Martens, S. Thutupalli, A. Fourri\'{e}re, and O. Hallastchek,
PNAS {\bf 110}, 10563 (2013).

\bibitem{Takahashi2010PNAS+Kveraga2011PNAS}
N. Takahashi {\it et al.}, PNAS {\bf 107}, 10244 (2010); K.
Kveraga {\it et al.}, PNAS {\bf 108}, 3389 (2011).

\bibitem{books}
A.S. Pikovsky, M. Rosenblum, and J. Kurths, {\em Synchronization:
A Universal Concept in Nonlinear Science}, Cambridge Nonlinear
Science Series (Cambridge University Press, Combridge, England,
2001);
G.V.Osipovsky, J. Kurths, and C. Zhou, {\em Synchronization in
Osillatory Networks}, Springer Series in Synergetics (Springer,
Berlin, 2007);
S. Boccaletti, {\em The Synchronized Dynamics of Complex Systems},
edited by A.C. Luo and G. Zaslavsky, Monograph Series on Nonlinear
Science and Complexity, Vol. 6 (Elsevier Science, Amsterdam,
2008).

\bibitem{Kuramoto}
Y. Kuramoto in {\em Proceedings of the International Symposium on
Mathematical Problems in Theoretical Physics}, Lecture Notes in
Physics, Vol. 39, edited by H. Araki (Springer-Verlag, Berlin,
1975);
{\it Chemical Oscillations, Waves, and Turbulence}
(Springer-Verlag, Berlin, 1984).

\bibitem{Winfree}
A.T. Winfree, J. Theor. Biol. {\bf 16}, 15~(1967);
{\it The Geometry of Biological Time} (Springer-Verlag, Berlin,
1980).

\bibitem{Kuramoto1984PTPS}
Y. Kuramoto, Prog. Theor. Phys. Suppl. {\bf 79}, 223~(1984).

\bibitem{Daido1987JPA}
H. Daido, J. Phys. A.:Math.Gen. {\bf 20}, L629 (1987).

\bibitem{Hong2004n2005PRE+2007PRL}
H. Hong, H. Park, and M.Y. Choi, Phys. Rev. E {\bf 70},
045204(R)~(2004);
{\em ibid.} {\bf 72}, 036217~(2005);
H. Hong, H. Chat{\'e}, H.
Park, and L.-H. Tang, Phys. Rev. Lett. {\bf 99}, 184101~(2007).

\bibitem{Son2010PRE}
S.-W. Son and H. Hong, Phys. Rev. E {\bf 81}, 061125~(2010).

\bibitem{Tang2011JSM}
L.-H. Tang, J. Stat. Mech.: Theor. Exp. P01034~(2011).

\bibitem{Hong2013PRE}
H. Hong, J. Um, and H. Park, Phys. Rev. E {\bf 87}, 042105~(2013).

\bibitem{Barabasi1995}
A.-L. Barab\'{a}si and H. E. Stanley, {\em Fractal Concepts of
Surface Growth} (Cambridge University Press, Cambridge, 1995).

\bibitem{Vicsek1984PRL}
T. Vicsek and F. Family, Phys. Rev. Lett. {\bf 52}, 1669~(1984).

\bibitem{Marro1999}
J. Marro and R. Dickman, {\em Nonequilibrium Phase Transitions in
Lattice Models} (Cambridge University Press, Cambridge, 1999).

\bibitem{Pazo2005PRE}
D. Paz{\'o}, Phys. Rev. E {\bf 72}, 046211~(2005).


\bibitem{Strogatz1991JSP}
S.H. Strogatz and R.E. Mirollo, J. Stat. Phys. {\bf 63},
613~(1991).

\bibitem{Strogatz1992PRL}
S.H. Strogatz, R.E. Mirollo, P.C. Matthews, Phys. Rev. Lett. {\bf
68}, 2730~(1992).

\bibitem{HP+Hong}
H. Park and H. Hong (private communication).

\bibitem{unpublished}
C. Choi, D. Kim, and M. Ha (unpublished data).

\bibitem{Zheng1999PRL}
B. Zheng, M. Schulz, and S. Trimper, Phys. Rev. Lett. {\bf 82},
1891~(1999).


\end{thebibliography}

\end{document}